\documentstyle[prl,aps,twocolumn,floats,psfig]{revtex}
\begin{document}
\noindent
{\bf Comment on ``Circumstantial Evidence for Critical Behavior in Peripheral
Au+Au Collisions at 35 MeV/nucleon.''}

\vspace{0.2cm}
Mastinu {\em et al.} recently reported the 
observation of several
positive signals possibly indicating critical behavior in peripheral collisions
of Au+Au at $E/A$=35 MeV \cite{Mas96}. 

In our comment, we 
examine the choice
of variables used to determine the presence (or absence) 
of critical behavior. We do this by
repeating the analysis of ref. \cite{Mas96} on ``data'' from a simulation 
with no critical behavior.

The simulation samples a charge distribution and
conserves charge (breaking up a source of size $Z_0$). 
The 
charge particle multiplicity $N_C$ is specified at
the outset. Within an event, at multiplicity 
$n$ (where $1\le n\le N_C-1$) the probability to emit a
particle of a given Z is 
\begin{equation}
P_n(Z)\propto 
e^{-\alpha Z}
\end{equation}
under the constraint that at each ``emission step'' $n$, 
the $Z$ of the emitted
particle be sufficiently small so that the event will satisfy the requirement
of containing $N_C$ particles. We chose $\alpha=0.3$ and $Z_0$ = 79.
The choice of exponential charge distribution (and $\alpha$=0.3) 
is arbitrary as is the specific
implementation of charge conservation. 

Using this simulation we constructed ``events'' and examined
the proposed observables for critical behavior.

\begin{figure}
\centerline{\psfig{file=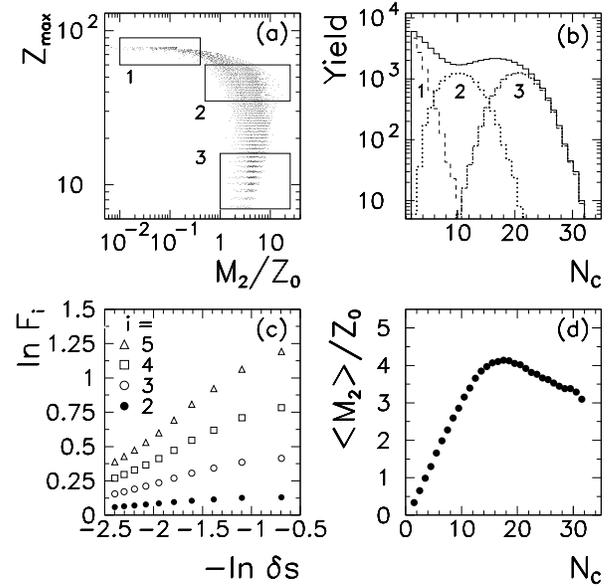,height=8.0cm}}
\caption{a) $Z_{max}$ vs $M_2/Z_0$ with cuts similar to those used in ref.
\protect\cite{Mas96}. b) Multiplicity
distribution of the simulation (solid line) and for cuts 1 (dashed), 2 (dotted) 
and 3 (dotted-dashed).
c) Log of the scaled factorial moments ($i$=2,3,4,5) 
as a function of
the negative log of the bin size $\delta s$ for cut 2 of ref.
\protect\cite{Mas96}.
d) Average $M_2/Z_0$ versus $N_C$.}
\label{4_panel}
\end{figure}

In Fig.\ref{4_panel}a is 
shown the Campi scatter plot of $Z_{max}$ versus $M_2/Z_0$. 
We observe the two-branch feature commonly interpreted as
indicating ``sub-critical'' and ``over-critical'' events.

By applying cuts similar to those in ref. \cite{Mas96}, we have plotted the
resulting multiplicity distribution (Fig.\ref{4_panel}b). Qualitative
agreement with the experimental data \cite{Mas96} 
is achieved with this simple event selection. We
question whether these cuts ``select'' events that can be associated with
critical behavior.

Using the same cut (2) in the Campi plot for ``potentially critical'' 
events as in ref. \cite{Mas96}, 
we have constructed the horizontally scaled
factorial moments. These moments are shown in 
Fig.\ref{4_panel}c. The linear rise with
decreasing bin size is quite apparent. 
It has already been pointed out \cite{Phair92,Cam95} 
that
spurious intermittency signals can be observed by mixing events of different
multiplicity, which is clearly the case for ref. \cite{Mas96} and for which the
authors appropriately express concern.

Finally, we show a plot of 
$M_2$ versus $N_c$ 
(Fig.\ref{4_panel}d). A
peak in such a plot is often mistakenly 
taken as an indication of critical behavior. While
we observe a peak, our simulation is one that 
assuredly contains no critical behavior. 
Perhaps one should instead examine the location,
height and width of the peak to search for evidence of critical
behavior. However, 
even then such an analysis may not reliably distinguish different
fragmentation mechanisms \cite{Cha95}.

We have repeated the analysis shown in Fig.~\ref{4_panel} 
for power law charge distributions and different
implementations of charge conservation but the qualitative results 
remain the same. 

Before doing this analysis, we were under the mistaken 
impression that the simple observables listed above 
give an indication of the presence or
absence of
critical behavior. Part of our confusion came from the vast commentary in the
literature that points to these observables as
indicators of critical phenomena. In fact, positive signals in all of these
observables are probably found in {\em any} simulation 
that conserves charge and where light particle emission
is preferred over heavy. 
And so we caution that the 
positive signals observed in ref.~\cite{Mas96} are
insufficient to establish critical behavior since they appear even in simple 
models which contain neither a phase transition nor critical behavior.

\vspace{0.1cm}
\noindent 
L. Phair, Th. Rubehn, L.G. Moretto, and G.J. Wozniak

Nuclear Science Division 

Lawrence Berkeley National Laboratory 

Berkeley, California 94720

\end{document}